\documentclass[a4paper,12pt]{article}
\usepackage{amsmath,amsfonts,amssymb,cite}
\newcommand{\be}{\begin{equation}}
\newcommand{\ee}{\end{equation}}

\begin{document}

\begin{center}
{\Large\bf Quark condensate and deviations from string-like behaviour of
meson spectra}
\end{center}

\begin{center}
{\large S.S.~Afonin} \\
V.A.~Fock Department of Theoretical Physics,
St.~Petersburg State University, Russia\\
e-mail: afonin24@mail.ru, Sergey.Afonin@pobox.spbu.ru
\end{center}

\begin{abstract}
I analyse the hypothesis that deviations from the linear meson mass
spectra appear due to the dynamical chiral symmetry breaking in
QCD. It is shown that the linear mass spectrum for the light,
non-strange vector and axial-vector mesons is then parametrized by
the constant $f_{\pi}$, being successful phenomenologically.
The toy model for deviations from linearity is proposed.
\end{abstract}

\noindent
Pacs: 12.38.Lg 

\section{Introduction}

It is widely believed that Quantum Chromodynamics and the boson
string theory are tightly connected. In the latter, the meson mass
spectrum $M^2(n)$ is linear with respect to the number of radial
excitation $n$. Phenomenological analysis of the experimental data
shows that this picture, at least qualitatively, indeed takes
place~\cite{ani}. Nevertheless, in the real world there are sizeable
deviations from linear trajectories. There arises the natural
question: what is the dynamics behind these deviations from the
string model? Partly, the deviations can be induced by mixing of
resonances and thresholds effects. Let us exclude these effects by
setting the widths of resonances to zero. In the QCD it is
equivalent to the limit $N_c\!=\!\infty$~\cite{hoof}. This limit is a
good approximation to the real world, having an accuracy of about
10\%. However, even in that limit one expects the linear mass
spectrum only at a large enough $n$, as it is in the
two-dimensional multicolor QCD --- the 't~Hooft model~\cite{dim2}.

In the present work I propose and analyze the hypothesis that, at
least in the large-$N_c$ limit, deviations from the linear
mass spectrum arise due to the dynamical chiral symmetry breaking
in QCD at low energies.

The convenient tools for the theoretical analysis are the two-point
correlation functions for the quark currents. In Euclidean space
there exist the Operator Product Expansion (OPE) for them at a large
euclidean momentum~\cite{svz}. Their OPE results in the set of
so-called chiral symmetry restoration (CSR) sum rules. Making use
of these sum rules for the case of light, non-strange vector (V) and
axial-vector (A)
mesons, I first analyze the consequences of the hypothesis for
the linear mass spectrum (sect.~2) and then consider a toy model
describing the corrections to the linear trajectories (sect.~3). I
conclude in sect.~4.

\section{Linear mass spectrum}

Let us consider the vector two-point functions which are defined
in Euclidean space as
\be
\Pi^J_{\mu\nu}=\int d^{4}e^{iQx}\left\langle\bar{q}\Gamma_{\mu}^J
q(x)\bar{q}\Gamma_{\nu}^Jq(0)\right\rangle=
\left(Q_{\mu}Q_{\nu}-\delta_{\mu\nu}Q^2\right)\Pi^J(Q^2).
\ee
There $\Gamma_{\mu}^V\!=\!\gamma_{\mu}$,
$\Gamma_{\mu}^A\!=\!\gamma_{\mu}\gamma_5$, and throughout the paper
$J\!=\!V\!,A$. In the chiral limit the axial-vector correlator
does not have the longitudinal part due to the pion pole.

In the chiral and large-$N_c$ limits, the OPE for the
objects under consideration is~\cite{svz}
\be
\label{ope}
\Pi^J(Q^2)=\frac{1}{4\pi^2}\left(1+\frac{\alpha_s}{\pi}\right)
\ln\!\frac{\Lambda^2}{Q^2}+\frac{\alpha_s}{12\pi}
\frac{\langle G^2\rangle}{Q^4}+\frac{4a^J}{9}\pi\alpha_s
\frac{\langle\bar{q}q\rangle^2}{Q^6}+\mathcal{O}\left(\frac{1}{Q^8}\right),
\ee
where $a^V\!=\!-7$, $a^A\!=\!11$, $\Lambda$ is a normalization
scale, $\langle G^2\rangle$ and $\langle\bar{q}q\rangle$ are the
gluon and quark condensate respectively. For simplicity,
the first-order perturbative correction to the partonic
logarithm will not be taken into account. As follows from Eq.~\eqref{ope},
the difference
$\Pi^V\!-\!\Pi^A$ (equal to zero in the perturbation theory) is
proportional to $\langle\bar{q}q\rangle^2/Q^6$. This reflects the fact
that the quark condensate is an order parameter of the chiral
symmetry breaking in QCD and that the chiral symmetry is restored
at high energies.

On the other hand, the functions $\Pi^J(Q^2)$ satisfy the
dispersion relation
\be
\Pi^J(Q^2)=\int_0^{\infty}\frac{ds}{s+Q^2}\frac{1}{\pi}Im\Pi^J(s)+
Subt.\,const.
\ee
If one assumes the string-like picture and the color confinement
in the large-$N_c$ QCD, the meson spectrum consists then of the
infinite set of equidistant narrow resonances with an universal
slope in all channels. These states saturate completely the
imaginary part of the correlation function:
\be
\label{sat}
\frac{1}{\pi}Im\Pi^J(s)=2f_{\pi}^2\delta(s)\delta_{A,J}+
2\sum_{n=0}^{\infty}F_J^2(n)\delta\left(s-M_J^2(n)\right).
\ee
In Eq.~\eqref{sat} $f_{\pi}$ is the pion decay constant,
$f_{\pi}\!\approx\!90$~MeV (in the chiral limit), $M_J(n)$ and $F_J$ are
the corresponding masses and decay constants, $\delta_{A,J}\!=\!1$ for $J\!=\!A$
and is zero otherwise. The first term in the l.h.s. of
Eq.~\eqref{sat} is due to the contribution of the pion pole to the
axial-vector correlator. If all quantum numbers are fixed the meson
spectrum has the linear parametrization in the string-like
picture:
\be
\label{spectr}
M_J^2(n)=m_J^2+\mu^2n,\qquad\quad F_J^2=const, \qquad n=0,1,\dots,
\ee
where $m_J^2$ is an intercept, $\mu^2$ is an universal slope, and
$n$ is the number of radial excitation.
According to the made hypothesis and the discussion above, corrections
to the spectrum~\eqref{spectr} are proportional to the quark
condensate. We omit them in the analysis of this section. Thus,
the two-point functions take the form
\be
\label{psi}
\Pi^J(Q^2)=\frac{2f_{\pi}^2}{Q^2}\delta_{A,J}+
\sum_{n=0}^{\infty}\frac{2F_J^2}{Q^2+m_J^2+\mu^2n}=
\frac{2f_{\pi}^2}{Q^2}\delta_{A,J}-\frac{2F_J^2}{\mu^2}\,\psi\!\left(
\frac{Q^2+m_J^2}{\mu^2}\right),
\ee
where the irrelevant (infinite) constants are cancelled. The $\psi$
function has an asymptotic representation at $z\!\gg\!1$
\be
\label{asymp}
\psi(z)=\ln{z}-\frac{1}{2z}-\sum_{n=1}^{\infty}\frac{B_{2n}}{2nz^{2n}},
\ee
there $B_{2n}$ denote the Bernoulli numbers.

Comparing expansion in $1/Q^2$~\eqref{asymp} for Eq.~\eqref{psi} with
the OPE~\eqref{ope}, one arrives at the following CSR sum rules:
\begin{align}
\label{0sr}
\frac{1}{8\pi^2}&=\frac{F_J^2}{\mu^2}\,, \\
\label{1sr}
f_{\pi}^2\delta_{A,J}&=F_J^2\left(\frac{m_J^2}{\mu^2}-\frac12\right)\,,\\
\label{2sr}
\frac{\alpha_s}{12\pi}\langle G^2\rangle&=
F_J^2\mu^2\left(\frac{m_J^4}{\mu^4}-\frac{m_J^2}{\mu^2}+\frac16\right)\,,\\
0&=-\frac23F_J^2\mu^4\left(\frac{m_J^6}{\mu^6}-\frac32\frac{m_J^4}{\mu^4}+
\frac12\frac{m_J^2}{\mu^2}\right)\,.
\label{3sr}
\end{align}
The l.h.s. of Eq.~\eqref{3sr} is set to zero by virtue of the made
hypothesis: the parameters of linear mass spectrum~\eqref{spectr}
do not depend on the quark condensate, which appears in the
l.h.s. of Eq.~\eqref{3sr} only after allowance for the corrections
to the spectrum~\eqref{spectr}. This factorization is crucial for
the given analysis.

Let us set $m_J^2=x\mu^2$. Eq.~\eqref{3sr} has the solutions:
$x_1=0$, $x_2=1/2$, and $x_3=1$. As follows from Eq.~\eqref{1sr},
the solution $x_2$ corresponds to the vector channel and for the
axial-vector case one has the only possibility --- the solution
$x_3$. The solution $x_1$ may be relevant to the pseudoscalar
channel which we do not consider here.

From Eqs.~\eqref{0sr} and~\eqref{1sr} one has for the decay
constants and slope\footnote{Here and further I do not trace the exact
$N_c$-dependence of the physical quantities. In particular, the factor
$\sqrt{3/N_c}$ for the slope $\mu$ is omitted in~\eqref{par}.}
\be
\label{par}
F_J=\sqrt{2}f_{\pi}\,, \qquad\quad \mu=4\pi f_{\pi}\,.
\ee
Thus, within this model, the whole V\!,A meson spectrum is
parametrized by only one constant $f_{\pi}$. Following the
classification of solutions given above, one obtains
($n=0,1,\dots$):
\be
\label{cool}
\begin{aligned}
M_V^2(n)&=16\pi^2f_{\pi}^2(\frac12+n),\\
M_A^2(n)&=16\pi^2f_{\pi}^2(1+n).
\end{aligned}
\ee
The two remarkable properties of the spectrum~\eqref{cool} should be
indicated: the value of the $\rho$ meson mass,
$M_{\rho}=2\sqrt{2}\,\pi f_{\pi}$, which is highly successful prediction
with respect to experiment~\cite{svz2}, and the natural
generalization of the Weinberg relation~\cite{weinberg}
($M_{a_1}=\sqrt{2}M_{\rho}$)
\be
M_{A}^2(n)=M_{V}^2(n)+M_{\rho}^2.
\ee

The comparison of model estimates and experimental data is
presented in Table~1. The data agree well within the large-$N_c$
approximation.

The ensuing from Eq.~\eqref{2sr} estimate for the value of gluon
condensate exceeds the corresponding phenomenological value,
$\frac{\alpha_s}{12\pi}\langle G^2\rangle=(360\,\text{MeV})^4$, in
the axial-vector channel,
$\frac{\alpha_s}{12\pi}\langle G^2\rangle=64\pi^2f_{\pi}^4=
(450\,\text{MeV})^4$, and gives a negative value in the vector
one. The difficulty with the gluon condensate in the sum rule~\eqref{2sr}
is known (see the first paper in~\cite{VA}). Within the scheme
under consideration this problem can be overcome by introducing
some non-linear corrections to the spectrum~\eqref{spectr}.

\section{Non-linear mass spectrum}

Let us consider the following ansatz for the meson mass spectrum:
\be
\label{spectr2}
M_J^2(n)=m_J^2+\mu^2n+\frac{f_{\pi}^2d_J}{(n+1)^{k_J}},\qquad
k_J>2.
\ee
By hypothesis, the last term in Eq.~\eqref{spectr2} (where the dimensional
parameter is singled out for clarity) represents a
decreasing in $n$ correction to the linear spectrum, caused by the
chiral symmetry breaking. Labeling this contribution as $\delta$,
one may wright (the index $J$ is omitted):
\be
\label{trans}
\sum_n\frac{F^2}{Q^2\!+\!m^2\!+\!\mu^2n\!+\!\delta}\!=\!
\sum_n\frac{F^2}{Q^2\!+\!m^2\!+\!\mu^2n}\!-\!
\sum_n\frac{F^2\delta}{(Q^2\!+\!m^2\!+\!\mu^2n)
(Q^2\!+\!m^2\!+\!\mu^2n\!+\!\delta)}.
\ee
The first term in the r.h.s. of Eq.~\eqref{trans} can be treated
as in the previous section. In the second one, the summation over~$n$
and the expansion in~$1/Q^2$ can be permuted by virtue of the
absolute convergence (up to $\mathcal{O}(1/Q^4)$).

Let us consider the axial-vector channel. The sum
rules~\eqref{0sr}, \eqref{1sr}, and~\eqref{3sr} are solved in the
zero order of the condensate expansion. Due to Eqs.~\eqref{ope}
and~\eqref{spectr2}, in the first order of this expansion
Eq.~\eqref{3sr} takes the form:
\be
d_A=\frac{11\pi\alpha_s}{8\pi^2\zeta(k_A-1)}
\frac{\langle\bar qq\rangle^2}{f_{\pi}^6}+
\mathcal{O}\left(\frac{\langle\bar
qq\rangle^4}{f_{\pi}^{12}}\right),
\ee
where $\zeta(x)$ is the Riemann zeta function. Let us impose the
condition of fulfillment of the sum rule~\eqref{2sr} in the first
order of the condensate expansion. This yields the equation for
the parameter $k_A$:
\be
\frac{1}{f_{\pi}^4}\frac{\alpha_s}{\pi}\langle
G^2\rangle=64\pi^2-48\zeta(k_A)d_A\,.
\ee

In the vector channel the corresponding equations look as follows:
\be
\begin{aligned}
d_V=-\frac{7\pi\alpha_s}{2\pi^2\left[2\zeta(k_V-1)-\zeta(k_V)\right]}
\frac{\langle\bar qq\rangle^2}{f_{\pi}^6}+
\mathcal{O}\left(\frac{\langle\bar
qq\rangle^4}{f_{\pi}^{12}}\right),\\
\frac{1}{f_{\pi}^4}\frac{\alpha_s}{\pi}\langle
G^2\rangle=-32\pi^2-48\zeta(k_V)d_V\,.
\end{aligned}
\ee

For the inputs $f_{\pi}=90$ MeV,
$\langle\bar qq\rangle=-(240\,\text{MeV})^3$,
$\frac{\alpha_s}{\pi}\langle G^2\rangle=(360\,\text{MeV})^4$, and
$\alpha_s=0.3$ the numerical calculations give $k_A\approx k_V\approx
2.1$ and show that the quark condensate terms indeed can be treated
as small parameters (their numerical contributions are less than
the large-$N_c$ counting). The presented solution lowers the mass of
$\rho$ meson by 40~MeV and enhances that of $a_1$ meson by 20~MeV.
The masses of excitations virtually do not change.

For the given fits, the correction to the Weinberg relation is
\be
M_{a_1}^2-2M_{\rho}^2\approx\frac{\langle\bar
qq\rangle^2}{20f_{\pi}^4},
\ee
that is, in this toy model the deviation from the Weinberg
relation represents an order parameter of the chiral symmetry
breaking in QCD.

\section{Summary}

In the present work I have considered the problem of matching the
vector and axial-vector two-point correlation functions in the
large-$N_c$ and chiral limits to their Operator Product Expansion at a
large euclidean momentum $Q^2$. In the first limit the two-point functions
are saturated by an infinite number of narrow resonances and,
given an ansatz for the meson spectrum, can be calculated exactly.
Expanding a correlator in powers of $Q^{-2}$ and comparing with
its OPE, one obtains the sum rule at each order of the expansion.
This activity was dealt with by various authors in the vector
channel~\cite{all}, in the vector and axial-vector channels on the
same footing~\cite{VA}, and in the vector, axial-vector, scalar,
and pseudoscalar channels simultaneously~\cite{we}.

The distinguishing
feature of the present analysis consists in the underlying hypothesis that
corrections to the string-like meson spectrum are related to the
chiral symmetry breaking in QCD. These contributions are
proportional to the quark condensate squared (in the first order of the
condensate expansion) and can be treated as small parameters. The
sum rule (s.r.) at $Q^{-6}$, which is sensitive to the chiral
symmetry breaking, gives then the set of solutions for the
intercepts (in the zero order of the condensate expansion). The s.r. at
$Q^{-2}$ allows to classify these solutions. The s.r. at $Q^{0}$
yields the residues (decay constants) and the s.r. at $Q^{-2}$ in
the axial-vector channel gives the value of universal slope.
As a result, the linear meson mass spectrum turns out to be
parametrized by only one constant $f_{\pi}$, being successful
phenomenologically. The s.r. at $Q^{-4}$ serves for checking the
solutions. It is found that this s.r. can not be satisfied without
non-linear corrections to the string-like spectrum. These
corrections turned out to be rather small, but they make the sum
rules self-consistent.

Thus, a simple toy-model is proposed, where the string-like part of
QCD spectrum and the part responsible for the chiral symmetry
breaking are naturally factorized. The phenomenological success of
the model may signify that the considered physical picture
is related to the real QCD.

\section*{Acknowledgments}

The author is grateful to A.A. Andrianov for useful discussions
and enlightening comments.
This work is supported by Grant RFBR 01-02-17152 and INTAS
Call 2000 Grant (Project 587).

\begin{center}
\begin{table}[b]
\begin{tabular}{|c|c|c|c|c|}
\hline
  $n$ & $1$ & $2$ & $3$ & $4$ \\
  \hline
  $M_V(n)$ & $800\,(769.8\pm0.8)$ & $1390\,(1465\pm25)$ & $1790\,(1700\pm20)$ & $2120\,(2149\pm17)$ \\
  \hline
  $M_A(n)$ & $1130\,(1230\pm40)$ & $1600\,(1640\pm40)$ & $1960$ & $2260$ \\
  \hline
  \hline
  \multicolumn{5}{|c|}{$\mu=1130\,(1090\div1140)$~\cite{ani},\qquad$F_{a_1}=130\,(123\pm25)$,\qquad$F_{\rho}=130\,(154\pm8)$} \\
  \hline
\end{tabular}
\caption{The linear meson mass spectrum (in MeV) for $f_{\pi}=90$ MeV.
The known experimental values~\cite{pdg} are displayed in brackets.}
\end{table}
\end{center}

\end{document}